\documentclass[%
 reprint, 
 amsmath,amssymb,
 aps,
 nofootinbib
 prb,
 superscriptaddress
 ]{revtex4-1}
\usepackage{scrextend}
\usepackage{graphicx}
\usepackage{dcolumn}
\usepackage{bm}
\usepackage{tabularx}
\usepackage[utf8]{inputenc}
\usepackage[english]{babel}
\usepackage[bottom]{footmisc}
\usepackage{color}

\usepackage{amsmath}
\usepackage{soul,xcolor}
\setstcolor{red}
\usepackage{hyperref}

\newcommand{\Lb}{\pazocal{L}}
\DeclareMathAlphabet{\pazocal}{OMS}{zplm}{m}{n}

\begin{document}

\preprint{APS/123-QED}

\title{First-principles study of disordered half-Heusler alloys 
\textit{X}Fe$_{0.5}$Ni$_{0.5}$Sn (\textit{X} = Nb, Ta) as thermoelectric prospects}

\author{Mohd Zeeshan}
\affiliation{Department of Physics, Indian Institute of Technology,
             Hauz Khas, New Delhi 110016, India}
\author{Chandan Kumar Vishwakarma}
\affiliation{Department of Physics, Indian Institute of Technology,
             Hauz Khas, New Delhi 110016, India}
\author{B. K. Mani}
\email{bkmani@physics.iitd.ac.in}
\affiliation{Department of Physics, Indian Institute of Technology,
Hauz Khas, New Delhi 110016, India}

\date{\today}
             
\begin{abstract} 

High lattice thermal conductivity in half-Heusler alloys has been 
the major bottleneck in thermoelectric applications. Disordered half-Heusler 
alloys could be a plausible alternative to this predicament. In this paper,
utilizing first-principles simulations, we have demonstrated the low lattice 
thermal conductivity in two such phases, NbFe$_{0.5}$Ni$_{0.5}$Sn and 
TaFe$_{0.5}$Ni$_{0.5}$Sn, in comparison to well-known half-Heusler alloy TiCoSb. 
We trace the low thermal conductivity to their short phonon lifetime, originating 
from the interaction among acoustic and low-lying optical phonons. We recommend 
nanostructuring as an effective route in further diminishing the lattice thermal 
conductivity. We further predict that these alloys can be best used in the 
temperature range 400-600~K and carrier concentration of less than 
10$^{21}$ carriers cm$^{-3}$. We found $\sim$35\% and $\sim$17\% enhancement 
in $ZT$ for NbFe$_{0.5}$Ni$_{0.5}$Sn and TaFe$_{0.5}$Ni$_{0.5}$Sn, respectively, 
as compared to TiCoSb. We are optimistic of the findings and believe these 
materials would attract experimental investigations.

\end{abstract}

\maketitle

\section{Introduction}





Thermoelectric materials are best known for their virtuosity to directly 
convert waste heat into electricity \cite{Jiang21}.  The figure of merit of such 
materials is expressed as $ZT=S^2\sigma T/\kappa$, where $S$ is the Seebeck 
coefficient, $\sigma$ is the electrical conductivity, and $\kappa (= \kappa_e + \kappa_l)$ 
is the thermal conductivity comprising the electronic and lattice parts \cite{Garmroudi21}.
Half-Heusler (hH) alloys have garnered a tremendous reputation as potential thermoelectric 
candidates in the past two decades owing to their robust properties such as thermal 
stability \cite{Roy12}, mechanical strength \cite{Berry17}, non-toxic elements \cite{Li16}, 
semiconducting nature of 18-valence electron count (VEC) compositions \cite{Zeeshan18}, 
and tunable band gaps \cite{Galanakis02}. These alloys exhibit magnificent electrical 
transport properties on account of multiple band valleys and band degeneracy \cite{Zhou18}. 
A record room temperature high power factor ($S^2 \sigma$) of 106 $\mu$W cm$^{-1}$ K$^{-2}$ 
is reported in this family in the case of \textit{p}-type Nb$_{0.95}$Ti$_{0.05}$FeSb \cite{He16}.
Albeit, the high lattice thermal conductivity originating from the relatively 
simple crystal structure of hH alloys overshadows their excellent electrical 
transport properties \cite{Chen13}. This is the foremost hurdle in maximizing the 
thermoelectric efficiency of hH alloys \cite{Xie12}. It is well-known that complex 
crystal structures having a large number of atoms exhibit remarkably low lattice 
thermal conductivity due to low velocity optical phonons \cite{Toberer11}.

Recently, based on similar guidelines, Anand {\em et al.}
devised a rather unique valence-balanced approach for minimizing
the lattice thermal conductivity in hH alloys \cite{Anand19}. The ordered
composition TiCoSb was manipulated by replacing Co with 50\%
Fe and 50\% Ni to arrive at the disordered composition
TiFe$_{0.5}$Ni$_{0.5}$Sb. This way the 18-VEC is preserved
and a relatively complex crystal structure is obtained
which could have low intrinsic lattice thermal conductivity.
To validate the point, TiFe$_{0.5}$Ni$_{0.5}$Sb was synthesized
and demonstrated to have almost threefold lower room temperature
lattice thermal conductivity than the parent composition TiCoSb.
This highlights the significance of the approach and suggests
that such disordered compositions could be preferred starting
crystal structures in the search of low lattice thermal conductivity in hH alloys. 
The research in such disordered hH alloys is relatively new
and is still in a growing phase. Interestingly, Anand {\em et al}. \cite{Anand19} 
have computationally shown that a much larger phase space
is possible for exploration.

Thus, immediate experimental investigations would interest the 
thermoelectrics community. Though, experiments without any theoretical 
inputs could be challenging. Theoretical Insights 
into the electronic structure, understanding of electrical and thermal 
transport, favorable dopants to improve the transport properties, and 
a comprehensive estimate of the figure of merit would greatly facilitate 
the experimental realization. 
With this as motivation, we select two parent hH alloys NbCoSn and TaCoSn 
to make disordered compositions in the search of low intrinsic lattice thermal 
conductivity. We selected cobalt-based compositions because cobalt reserves 
are surmised as critical due to their extensive usage and it would be beneficial 
to replace Co with earth-abundant elements \cite{Liu19}. Thus, we replace Co 
with 50\% Fe and 50\% Ni in NbCoSn and TaCoSn to arrive at NbFe$_{0.5}$Ni$_{0.5}$Sn 
and TaFe$_{0.5}$Ni$_{0.5}$Sn, respectively. 

In this paper, utilizing the first-principles simulations, we have explored the
thermoelectric potential of disordered hH alloys NbFe$_{0.5}$Ni$_{0.5}$Sn (NFNS) 
and TaFe$_{0.5}$Ni$_{0.5}$Sn (TFNS). 
Throughout this paper, we have used the well-known thermoelectric material TiCoSb 
of hH family as a reference to compare our results. 
Specifically, we have tried to address the following
questions: 
i) Can the proposed alloys offer a lower lattice thermal 
conductivity than TiCoSb? If yes, what could be the origin of 
such reduction?  
ii) What would be the impact on electronic structure and 
transport properties on replacing cobalt with iron and nickel, 
iii) And finally, can these disordered hH alloys lead to an enhanced figure 
of merit. If yes, at what temperature and carrier concentration range. 

The paper is organized as follows: In Sec. II, we have provided a brief 
description of the computational methods employed in the calculations. 
In Sec. III, we present and discuss our results on electrical and thermal 
transport properties and based on that we estimate the figure of merit 
for the proposed disordered hH alloys. 
And finally, the summary of the paper is presented in Sec. IV.

Throughout this paper, we have used TiCoSb as a reference
to compare our calculated results. This is because TiCoSb is a well-known 
thermoelectric material in hH family. For convenience, we call NbFe$_{0.5}$Ni$_{0.5}$Sn 
and TaFe$_{0.5}$Ni$_{0.5}$Sn as NFNS and TFNS, respectively.

\section{Computational Methodology}

We performed the density functional theory based first-principles simulations
using Vienna \textit{ab initio} simulation package
(VASP) \cite{Kresse96, KressePRB} to examine the thermoelectric prospects
of TiCoSb, NFNS and TFNS systems. The generalized-gradient approximation based
Perdew-Burke-Ernzerhof (PBE) \cite{Perdew96} pseudopotential was employed to
account for the exchange-correlation among electrons.  The plane wave bases with an
energy cutoff of 500 eV were used in all the calculations.  Crystal structures
were optimized using full relaxation calculations with the help of
conjugate gradient algorithm using Monkhorst-Pack $k$-mesh of 11$\times$11$\times$11.
An energy convergence criterion of 10$^{-8}$ eV was used for all
self-consistent-field (SCF) calculations, and atomic forces were optimized
up to 10$^{-7}$ eV/{\AA}. The SCF calculations for energy were carried out
using a denser $k$-mesh of 21$\times$21$\times$21.

Phonon dispersions and lattice thermal conductivity were computed using a
2$\times$2$\times$2 (96 atoms) supercell. For phonons, we used the finite
displacement approach as implemented in PHONOPY \cite{Togo15}.
Force constants were calculated using the atomic displacement of 0.01 \AA\; 
and $\Gamma$-centered $k$-mesh of 2$\times$2$\times$2. For lattice thermal 
conductivity, the second- and 
third-order force constants were computed using the atomic displacements of 0.03 \AA. 
We evaluated 2726 displacements to extract the third-order force constants.
The lattice thermal conductivity was obtained by solving the linearized phonon
Boltzmann transport equation within the single-mode relaxation time approximation
as implemented in PHONO3PY \cite{TogoPRB}. The values of lattice thermal conductivity
were tested for convergence with respect to $q$-mesh size and it was found to converge
well with the $q$-mesh of 21$\times$21$\times$21.

Electron lifetimes and electrical transport coefficients were obtained
by solving the Boltzmann transport equation as implemented in AMSET \cite{Ganose21}.
We tested the convergence of transport properties with respect to increasing $k$-grid.
We found that augmenting $k$-grid beyond 21$\times$21$\times$ 21 has negligible
change in the transport properties. Further, the transport properties were tested
for the convergence in terms of interpolation factor. A default interpolation factor
of 10 lead to the converged transport properties.

\begin{figure}
\centering\includegraphics[scale=0.22]{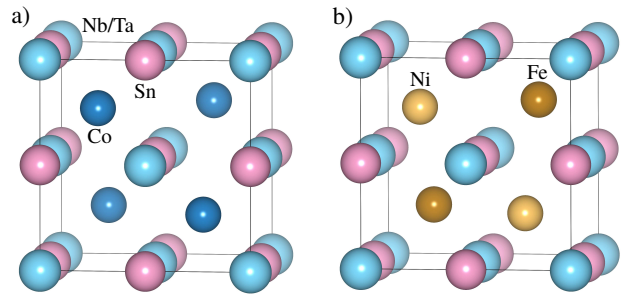}
\caption{Crystal structure of (a) \textit{X}CoSn
and (b) \textit{X}Fe$_{0.5}$Ni$_{0.5}$Sn, where \textit{X}
= Nb/Ta. The structures are produced by using VESTA \cite{Momma11}.}
\label{Crystal}
\end{figure}

\section{Results and Discussion}

\subsection{Crystal Structure and Phonon Dispersions}

The crystal structure of hH alloys, \textit{XYZ},
exists in \textit{F$\bar{4}$3m} symmetry, where
\textit{X} and \textit{Z} form the zinc-blende
type sublattice and \textit{Y}-atoms occupy half
of the tetrahedral voids \cite{Casper08}. To arrive at the aforementioned 
compositions of the disordered alloys, we replaced Co 
in \textit{X}CoSn (\textit{X} = Nb, Ta) with 50\% Fe and 50\% Ni. 
Since all the tetrahedral voids positions are invariant 
under \textit{F$\bar{4}$3m} symmetry, 
the Fe/Ni can be introduced at any position. Thuswise,
we obtained NbFe$_{0.5}$Ni$_{0.5}$Sn and TaFe$_{0.5}$Ni$_{0.5}$Sn structures,
as shown in Fig. \ref{Crystal}. However, the symmetry of these structures 
is expectedly reduces from \textit{F$\bar{4}$3m} to \textit{P$\bar{4}$m2}.
For optimization, we allowed all key degrees freedoms such as atomic positions, 
cell shape, and cell volume to relax, leading to an accurate ground state 
crystal structure. The optimized parameters for parent crystal structure, 
TiCoSb, are listed in Table~\ref{Opt}.  We obtained a quite good agreement with
the reported experimental and theoretical data. The calculated and experimental lattice
constants \cite{Sekimoto06} are consistent whereas the calculated band gap is
in fairly good agreement with previous calculations \cite{Xu12, Wang09, Zeeshan17}.
The deviation of the calculated band gap from the
experimental value could be attributed to the 
non-stoichiometric composition and antisite
disordering as suggested by Sekimoto \textit{et al} \cite{Sekimoto05}. 
Nonetheless, it is important to note that the band gap
survives on going from \textit{X}CoSn to 
\textit{X}Fe$_{0.5}$Ni$_{0.5}$Sn (\textit{X} = Nb, Ta),
which is a good indication for a thermoelectric material.

\begin{table}[]
\caption{Cell parameters and band gaps from our calculations. The 
	experimental and theoretical values are mentioned in parentheses 
	and square brackets, respectively.}
\centering
\begin{tabular*}{\columnwidth}{l @{\extracolsep{\fill}} cccc}
\hline 
\hline
System		                & a (\AA)               & c (\AA)       &E$_g$ (eV) \\
\hline
	TiCoSb                      &5.88 (5.88)        &    &1.05 (0.19) \cite{Sekimoto06} \\
			    &                       &             &[1.06] \cite{Xu12}\\
			    &                       &             &[1.06] \cite{Wang09}\\
NbFe$_{0.5}$Ni$_{0.5}$Sn    &5.97                   &6.00         &0.10   \\            
TaFe$_{0.5}$Ni$_{0.5}$Sn    &5.97                   &5.99         &0.16   \\
\hline
                            &      &Wyckoff Positions               &       \\ \hline
Ti                          &0.00                   &0.00           &0.00   \\                            
Co                          &0.25                   &0.25           &0.25  \\                            
Sb                          &0.50                   &0.50           &0.50   \\ \hline
Nb/Ta                       &0.00                   &0.00           &0.98   \\
Fe                          &0.25                   &0.25           &0.75   \\ 
Ni                          &0.25                   &0.25           &0.25   \\ 
Sn                          &0.50                   &0.50           &0.49   \\
\hline
\hline
\end{tabular*}
\label{Opt}
\end{table}

\begin{figure*}
\centering\includegraphics[scale=0.33]{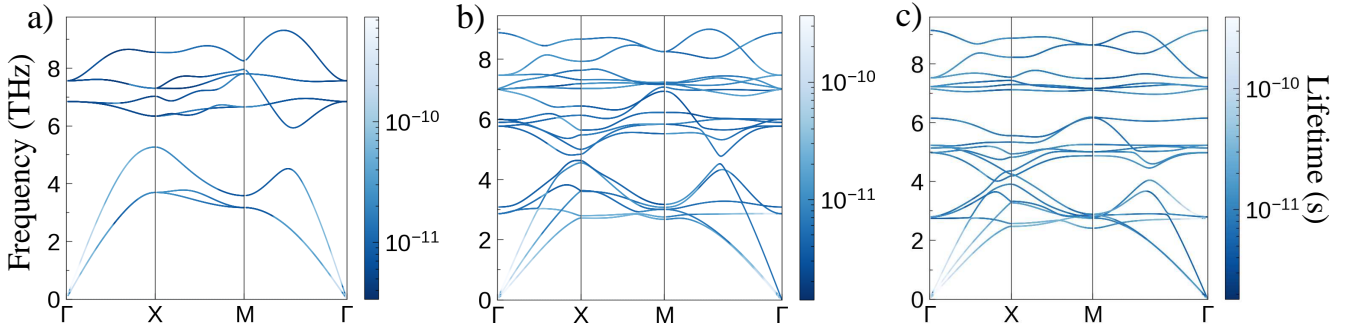}
\caption{Calculated phonon dispersions of (a) TiCoSb, (b) NbFe$_{0.5}$Ni$_{0.5}$Sn, 
	and (c) TaFe$_{0.5}$Ni$_{0.5}$Sn. The colorbar indicates the intensity of 
	phonon lifetime. The figures are plotted courtesy of ThermoPlotter \cite{TP}.}
\label{Phonons}
\end{figure*}

Using the optimized ground state structure, we computed the phonon dispersions 
by solving the eigenvalue equation (as implemented in the Phonopy \cite{Togo15})
\begin{equation}
  \sum_{\beta\tau'} D^{\alpha\beta}_{\tau \tau'}
        (\mathbf{q}) \gamma^{\beta\tau'}_{\mathbf{q}j} = 
        \omega^2_{\mathbf{q}j}\gamma^{\alpha\tau}_{\mathbf{q}j}. 
\end{equation}
Here, the indices $\tau, \tau'$ represent the atoms,
$\alpha, \beta$ are the Cartesian coordinates, ${\mathbf{q}}$ is
a wave vector and $j$ is a band index.  $D(\mathbf{q})$ refers to as
the dynamical matrix, and $\omega$ and $\gamma$ are the corresponding
phonon frequency and polarization vector, respectively.
Figure~\ref{Phonons} shows the phonon dispersions for NFNS and TFNS, 
and the reference material TiCoSb. The absence of imaginary phonon frequencies 
suggests the dynamical stability of the proposed materials.  Interestingly, these 
materials are also predicted to be thermodynamically stable \cite{Anand19}. 
Taken together, this greatly increases the likelihood of synthesis of NFNS and 
TFNS.  Both NFNS and TFNS exhibit similar topology of phonons, which could be 
attributed to the same crystal structure. As to be expected, the phonon 
density of states show that the acoustic modes are predominantly occupied by 
the heavy atoms (Sb in TiCoSb, Sn in NFNS, and Ta in TFNS), high frequency 
optical modes by lighter atom Fe, and mid-frequency region collectively 
by the rest atoms.

As discernible from the figure, there is an apparent gap between the acoustic 
and optical (\textit{a-o}) phonon modes for TiCoSb. Such a gap is often observed 
in hH alloys and is associated with their high lattice thermal conductivity \cite{Yu21}. 
Interestingly, this \textit{a-o} gap disappears in NFNS and TFNS which opens 
up the possibility of phonon-phonon scattering among acoustic and low-lying optical 
modes. Thereby, the acoustic phonon modes, predominantly responsible for lattice 
thermal conductivity, are likely to have a small lifetimes. To validate the same, 
we calculated the phonon lifetime at 300~K and the data from this is shown by 
the color bar in Fig.~\ref{Phonons}. Clearly, the phonon lifetime of NFNS and 
TFNS are shorter than that of TiCoSb, which hints at their low lattice thermal 
conductivity in comparison to TiCoSb. Thus, it will be interesting to examine 
the lattice thermal conductivity trend of the proposed materials.

\subsection{Lattice Thermal Conductivity}

\begin{figure}
\centering\includegraphics[scale=0.4]{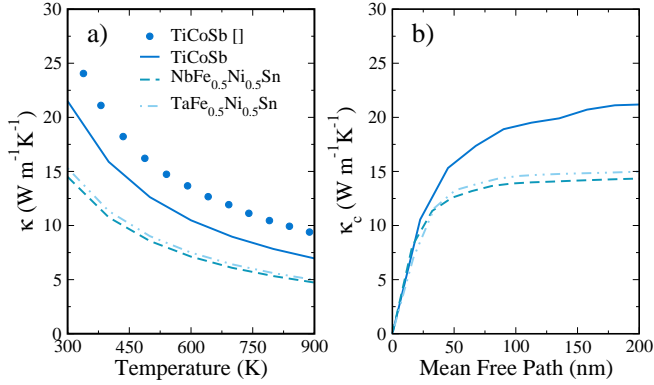}
\caption{Calculated (a) average lattice thermal conductivity
as a function of temperature and (b) cumulative lattice
thermal conductivity as a function of phonon mean free
path for TiCoSb, NbFe$_{0.5}$Ni$_{0.5}$Sn, and
TaFe$_{0.5}$Ni$_{0.5}$Sn. 
The empty circles represent the experimental data for 
TiCoSb \cite{Sekimoto06}.}
\label{Kappa}
\end{figure}

\begin{figure*}
\centering\includegraphics[scale=0.50]{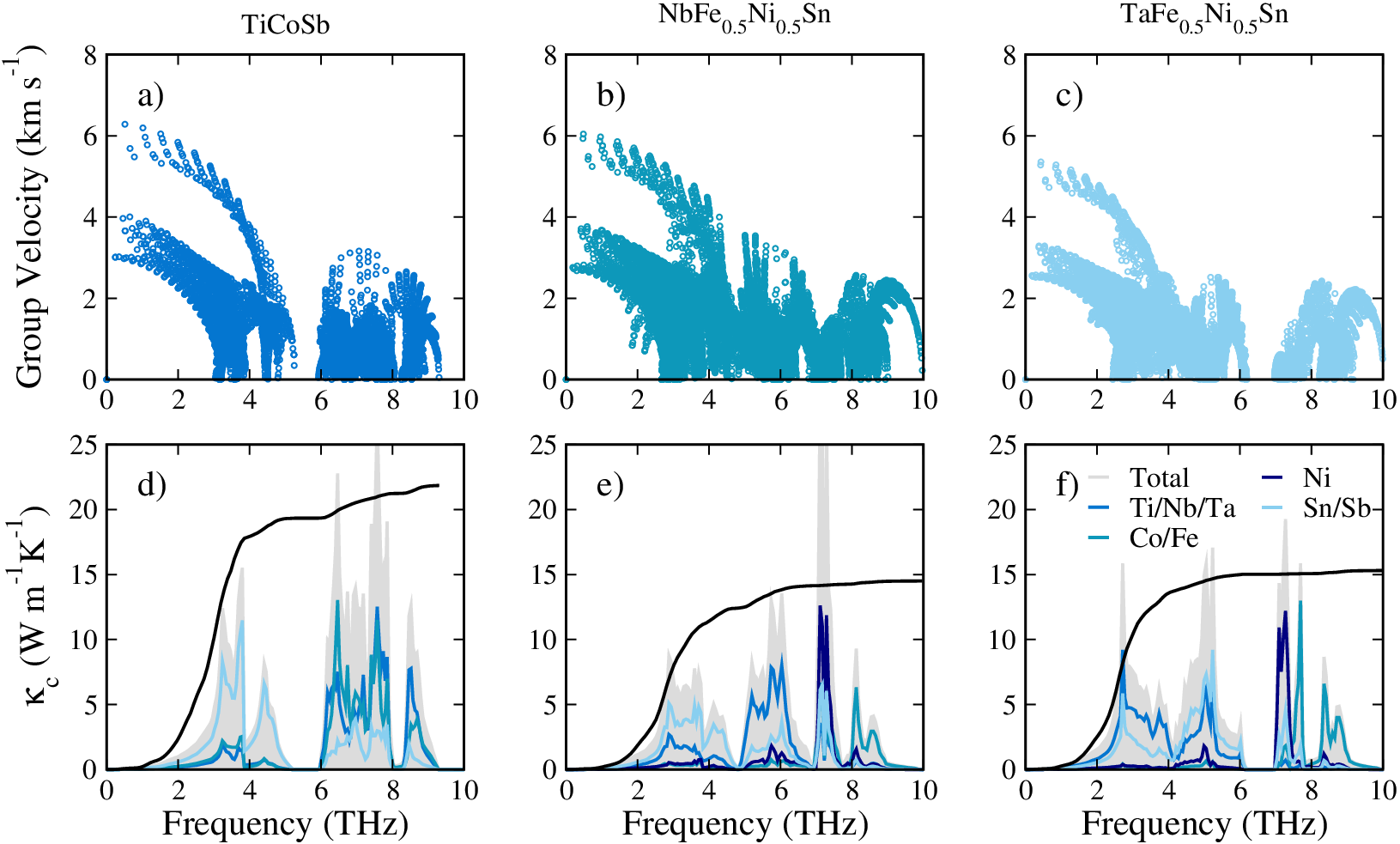}
\caption{Calculated [(a), (b), and (c)] average phonon group velocities 
	and [(d), (e), and (f)] cumulative lattice thermal conductivity at 300~K 
	for TiCoSb, NbFe$_{0.5}$Ni$_{0.5}$Sn, and TaFe$_{0.5}$Ni$_{0.5}$Sn, 
	respectively, as a function of phonon frequencies. The phonon density 
	of states are superimposed in arbitrary units.}
\label{Gv}
\end{figure*}

Next we examine the lattice thermal conductivity, $\kappa_L$, for the 
proposed disordered alloys. $\kappa_L$ is obtained by solving the linearized 
phonon Boltzmann transport equation as implemented in the PHONO3PY \cite{TogoPRB}. 
Figure~\ref{Kappa}(a) shows the calculated $\kappa_L$ 
for NFNS, TFNS and the reference material TiCoSb. For comparison, we have also provided
the experimental values for TiCoSb. We obtained slightly underestimated $\kappa_L$ 
for TiCoSb in comparison to experimental values. It is to be, however, emphasized 
that the trend as function of temperature is same for both theory and experiment. 
The $\kappa_L$ values of NFNS and TFNS bear close resemblance throughout the 
temperature range. The $\kappa_L$ for TFNS is slightly higher than that of 
NFNS at 300~K, but later on the two data approach each other and the difference 
is negligible at 900~K. Interestingly, the $\kappa_L$ of NFNS and TFNS is 
significantly lower than TiCoSb at 300~K, highlighting the importance of 
the substitution of Co with 50\% Fe and Ni. Beyond 300~K, the difference 
in $\kappa_L$ of the proposed materials and TiCoSb decreases. And, the reason 
for this could be attributed to the increasing phonon-phonon scattering 
with temperature.
The smaller $\kappa_L$ for NFNS and TFNS in comparison to 
TiCoSb could be attributed to their small phonon lifetimes, arising from 
the interaction among acoustic and low-lying optical modes. To gain further insight 
into the behavior of $\kappa_L$, we look into the phonon group velocities. 
As discernible from Figs.~\ref{Gv}(a) - (c), the magnitude of group velocities 
is not significantly different in all three systems. This suggests that the 
phonon group velocities may not have played an important role in reducing 
the $\kappa_L$ of the proposed materials.

Now we turn to understand the contribution of individual atoms to lattice 
thermal conductivity.  For this, we calculated the cumulative $\kappa_L$ as 
a function of phonon frequencies, shown in Figs.~\ref{Gv}(d) - (f).
The major contribution to $\kappa_L$ is observed from the modes below 4 THz, 
around 81\% in TiCoSb, 77\% in NFNS, and 88\% in TFNS. These modes are dominated 
by Sb in TiCoSb, Sn in NFNS, and Ta in TFNS. Thus, further reduction in 
$\kappa_L$ may be facilitated by doping at Sn-site in NFNS, and Ta-site in TFNS. 
Further, we analyzed the cumulative $\kappa_L$ with respect to phonon mean 
free path at 300~K and the data from this is shown in the, Fig.~\ref{Kappa}(b).
As we observe from the figure, the phonons majorly contributing to $\kappa_L$ 
have a mean free path in the range 0-50 nm in case of NFNS and TFNS. However, 
the mean free path ranges up to 90 nm for phonons boosting $\kappa_L$ in TiCoSb. 
This shows that the phonon mean free path of NFNS and TFNS are shorter 
as compared to TiCoSb, which is an indicative of low $\kappa_L$ in NFNS/TFNS. 
This short mean free path is consistent with the small phonon lifetime of NFNS/TFNS, as 
discussed before. It is noteworthy that $\sim$15\% contribution to $\kappa_L$ 
in NFNS/TFNS comes from the phonons with a mean free path larger than 50 nm.
Thus, in our opinion, nanostructuring could be an effective strategy for 
further diminishing the $\kappa_L$ for the proposed materials.

Altogether, we have demonstrated the low $\kappa_L$
in disordered hH alloys NFNS and TFNS in comparison
to well-known hH alloy TiCoSb. In the next section,
we explore the electronic structure and electrical
transport properties of the proposed materials.

\begin{figure*}
\centering\includegraphics[scale=0.6]{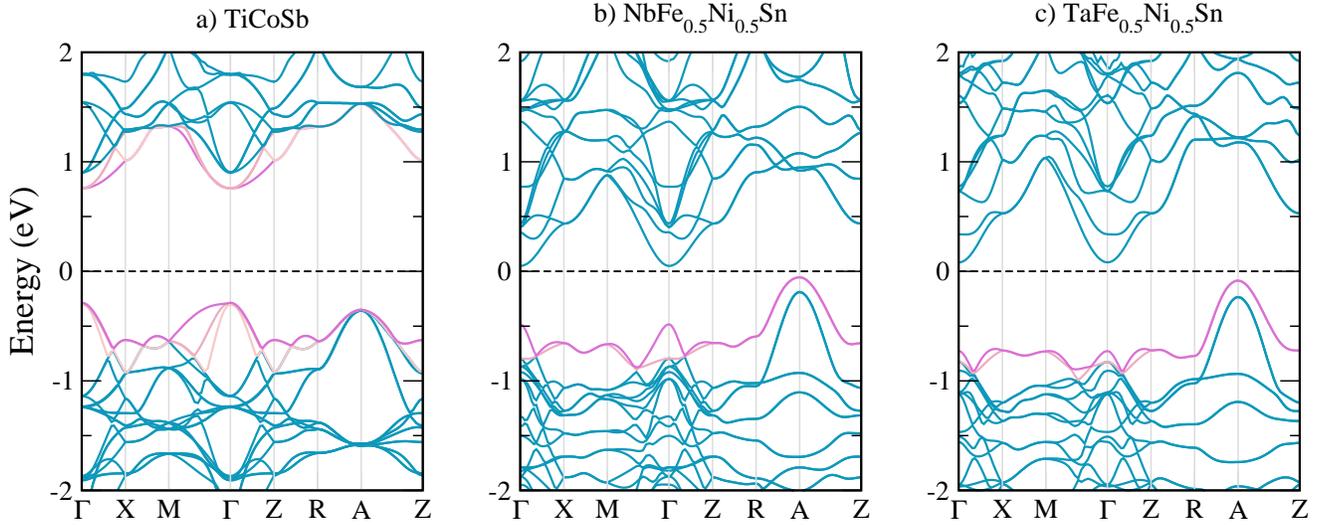}
\caption{Calculated electronic band structures for (a) TiCoSb, 
	(b) NbFe$_{0.5}$Ni$_{0.5}$Sn, and (c) TaFe$_{0.5}$Ni$_{0.5}$Sn. 
	The degenerate bands at VBM and CBM are highlighted in 
	different colors.}
\label{Bands}
\end{figure*}

\begin{figure*}
\centering\includegraphics[scale=0.5]{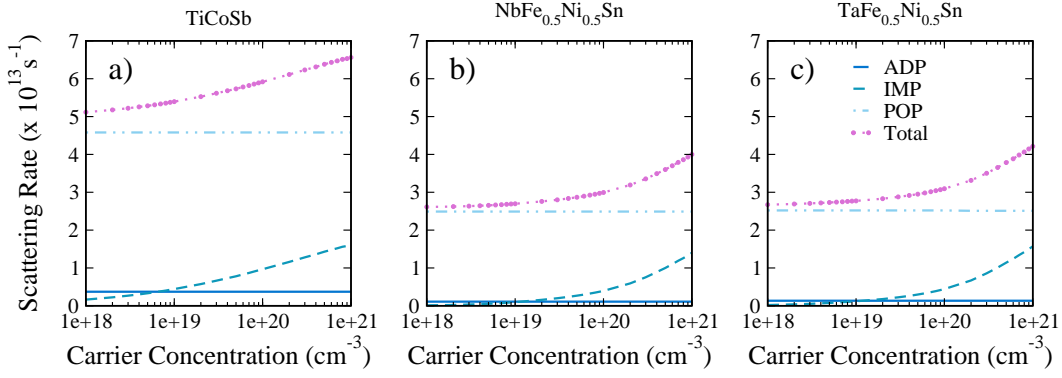}
\caption{Calculated scattering rates at 300~K for
\textit{p}-type (a) TiCoSb, (b) NbFe$_{0.5}$Ni$_{0.5}$Sn,
and (c) TaFe$_{0.5}$Ni$_{0.5}$Sn systems.}
\label{RT}
\end{figure*}

\subsection{Electronic Structure and Transport Properties}

\begin{figure*}
\centering\includegraphics[scale=0.31]{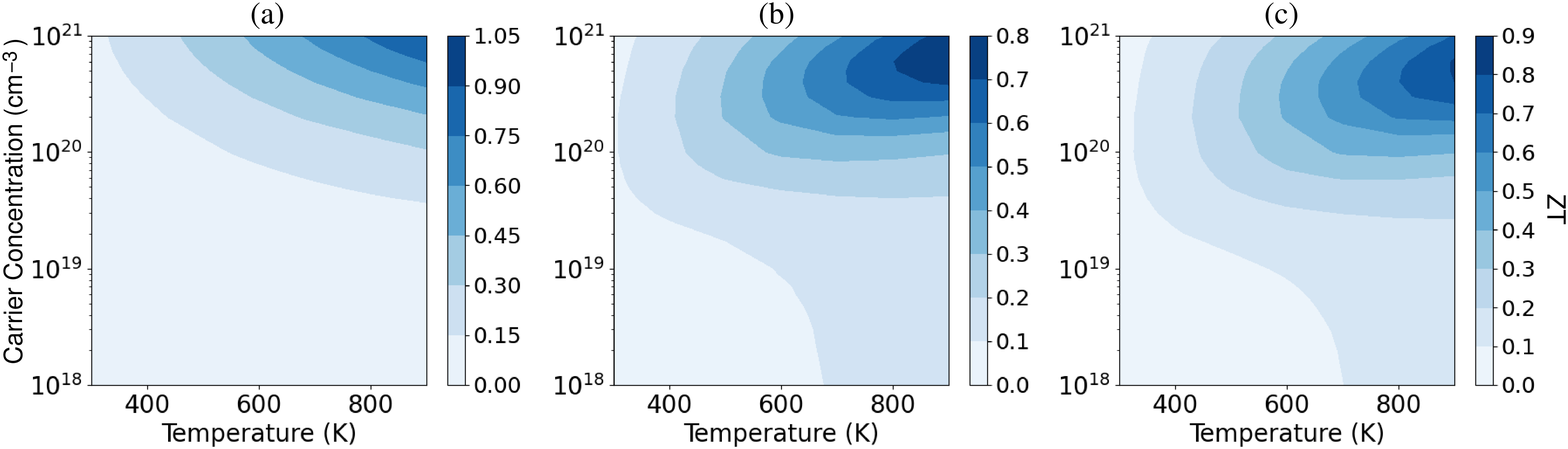}
\caption{Calculated figure of merit as a function
of temperature and carrier concentration for (a) TiCoSb, 
	(b) NbFe$_{0.5}$Ni$_{0.5}$Sn, and (c) TaFe$_{0.5}$Ni$_{0.5}$Sn 
	systems.}
\label{ZT}
\end{figure*}

The calculated electronic band structures for NFNS, TFNS, and the reference 
material TiCoSb is shown in Fig.~\ref{Bands}. The electronic structure 
of TiCoSb is well explored in the literature \cite{Xu12, Wang09, Zeeshan17}.
Our computed data is consistent with the literature in terms of reported 
theoretical band gap of 1.06 eV \cite{Xu12, Wang09}.
As mentioned earlier, both NFNS and TFNS are observed to be semiconducting. 
The nature of the band gap is direct at the $\Gamma$-position in the case of 
TiCoSb, whereas it is indirect (A-$\Gamma$) for both NFNS and TFNS. The band 
degeneracy, as observed in the case of TiCoSb, is a characteristic of exemplary 
electrical transport properties \cite{Tang15}. The valence band maximum (VBM) and
conduction band minimum (CBM) are threefold degenerate in the case of TiCoSb. 
However, the degeneracy is reduced to twofold at VBM in NFNS and TFNS. Not only 
this, the band gap is drastically reduced to 0.1 eV. We trace this reduced band 
gap majorly to the states originating from Fe-atoms. This can can be seen from 
the atom projected density of states given in the supplemental material \cite{Suppl}. 
The reduced band degeneracy and band gap are likely to affect the electrical 
transport properties. We anticipate a better Seebeck coefficient in TiCoSb on 
account of degenerate bands, whereas a better electrical conductivity in NFNS 
and TFNS because of the small band gap.

\begin{figure}[ht]
\centering\includegraphics[scale=0.4]{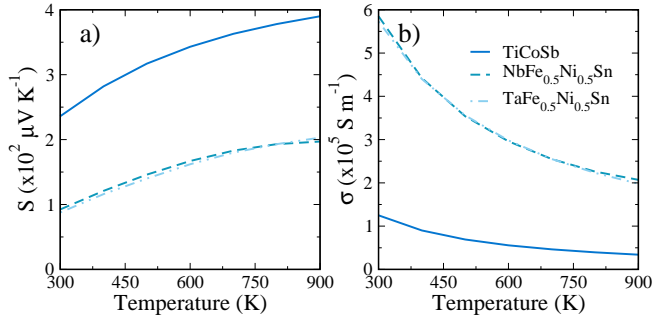}
\caption{Calculated (a) Seebeck coefficient and
(b) electrical conductivity as a function of
temperature at 4$\times$10$^{20}$ carriers
cm$^{-3}$ for TiCoSb, NbFe$_{0.5}$Ni$_{0.5}$Sn, and 
TaFe$_{0.5}$Ni$_{0.5}$Sn systems.}
\label{PF}
\end{figure}

\begin{table}
\caption{Calculated figure of merit for TiCoSb, NbFe$_{0.5}$Ni$_{0.5}$Sn, and 
	TaFe$_{0.5}$Ni$_{0.5}$Sn at different temperatures and carrier 
	concentrations.}
\centering
\begin{tabular*}{\columnwidth}{c @{\extracolsep{\fill}} ccccc}
\hline 
\hline
&               & &\multicolumn{3}{c}{\textit{ZT}}   \\ \cline{4-6}
&\textit{T} (K) &\textit{n} (cm$^{-3}$)    &TiCoSb &NFNS   &TFNS   \\
\hline
&300            &1$\times$10$^{20}$        &0.05    &0.09   &0.08   \\
&400            &2$\times$10$^{20}$        &0.09    &0.14   &0.12   \\
&500            &3$\times$10$^{20}$        &0.13    &0.18   &0.16   \\
&600            &4$\times$10$^{20}$        &0.17    &0.23   &0.20   \\
&700            &5$\times$10$^{20}$        &0.21    &0.25   &0.23   \\
&800            &6$\times$10$^{20}$        &0.24    &0.25   &0.26   \\
&900            &7$\times$10$^{20}$        &0.26    &0.26   &0.26   \\
\hline
\hline
\end{tabular*}
\label{FOM}
\end{table}

To get further insight into the transport properties, next we computed the 
electrical transport coefficients $S$, $\sigma$, and $\kappa_e$. These are 
derived from the generalized transport coefficient (as implemented in 
AMSET \cite{Ganose21}), expressed as
\begin{equation}
\Lb^n_{\alpha \beta} = e^2 \int{\textstyle \sum_{\alpha \beta}
        (\varepsilon) (\varepsilon - \varepsilon_F)^n
        \left[-\frac {\partial f^0}{\partial \varepsilon} \right]} d\varepsilon,
\end{equation}
where $\varepsilon_F$ is Fermi level at a particular doping,
$f^0$ is the Fermi distribution function,
and $\sum_{\alpha \beta} (\varepsilon)$ is the spectral
conductivity.
Since electrical transport parameters are susceptible to the charge carriers' 
relaxation time, it is important to evaluate the scattering rate for accurately 
governing the transport properties. 
We obtained the same using scattering mechanisms such as acoustic
deformation potential (ADP), ionized impurity (IMP), polar optical phonon (POP), and
the piezoelectric (PZ) scattering. The various parameters computed for different
scattering mechanisms include deformation potential, elastic constant, piezoelectric
coefficient, polar optical phonon frequency, and static and high-frequency 
dielectric constants.

The calculated scattering rates at different temperatures and carrier 
concentrations are provided in the supplemental material \cite{Suppl}. Here, we 
have shown the calculated scattering rates at 300~K for TiCoSb, NFNS, and 
TFNS, Fig.~\ref{RT}. 
For clarity, we choose to show the results only for \textit{p}-type carrier 
concentration. This is because we obtained significant results for \textit{p}-type 
in comparison to \textit{n}-type carrier concentration. Further, TiCoSb is well-known 
as a \textit{p}-type thermoelectric material. Nonetheless, as discernible from the 
figure, the contribution of ADP and POP scattering at 300~K remains constant 
throughout the considered range of carrier concentration. The IMP scattering 
however increases gradually and then gains upon ADP beyond carrier concentration 
of 10$^{19}$ carriers cm$^{-3}$. The increase in IMP with increasing carrier 
concentration could be attributed to the increased ionized impurities.
However, the dominant contribution is from POP in all the cases. The total 
scattering rate mirrors the trend of IMP scattering. Overall, at 300~K, the 
scattering rate ranges 5 -- 6.5$\times$10$^{13}$ s$^{-1}$ for TiCoSb and 
2.6 -- 4.2$\times$10$^{13}$ s$^{-1}$ for NFNS and TFNS. Incorporating these 
scattering rates we computed the figure of merit and the data from this is 
shown in Fig.~\ref{ZT}. It is to be however mentioned that, we have used 
$\kappa_L$ of parent compositions for evaluating the figure of merit since 
it is computationally challenging to calculate $\kappa_L$ at each 
carrier concentration.

The trend of the figure of merit is quite obvious from the trend of its 
components $S$, $\sigma$, and $\kappa_e$ given in the supplemental 
material \cite{Suppl}. As to be expected, the $S$ and $\sigma$ show the 
opposite trends, i.e, $S$ is maximum near low carrier concentrations, 
whereas $\sigma$ exhibits higher values at high carrier concentrations.
The $S$ dominates in case of TiCoSb, whereas $\sigma$ in NFNS and TFNS 
is almost fourfold higher in comparison to TiCoSb. This is consistent with 
the observation made from the electronic structure. It is interesting to note
that the power factor of all three systems is of the same order. The maximum 
power factor obtained is 105 $\mu$W cm$^{-1}$ K$^{-2}$. Thus, $\kappa_e$ 
is quite crucial which is significantly high in NFNS and TFNS.

Taking all together, the highest figure of merit obtained are 1.0, 0.8,
and 0.9, respectively, for TiCoSb, NFNS, and TFNS
at 900 K and carrier concentration of 1$\times$10$^{21}$
carriers cm$^{-3}$. 
This suggest that the proposed alloys at high temperatures and high 
carrier concentration range are competitive with TiCoSb. At lower 
temperatures and lower concentrations, however, the proposed alloys show
an edge over the parent TiCoSb.
The figure of merit for TiCoSb is less than 0.15 up to 500~K in the 
carrier concentration range 1$\times$10$^{18}$ to 1$\times$10$^{20}$ 
carriers cm$^{-3}$. On the other hand, the $ZT$ of NFNS and TFNS 
increases consistently with temperature and carrier concentration.
As a matter of fact, NFNS and TFNS have improved $ZT$ value than TiCoSb 
in the temperature region 300-700~K for the carrier concentration 
in the range 1$\times$10$^{18}$--5$\times$10$^{20}$ carriers cm$^{-3}$. 
As evident from the Table~\ref{FOM}, for a fixed carrier concentration 
of 4$\times$10$^{20}$ carriers cm$^{-3}$, at 600~K, the $ZT$ for NFNS 
and TFNS are larger by $\sim$35\% and $\sim$17\%, respectively, than TiCoSb. 
This shows that the proposed materials can be better thermoelectric
prospects as compared to TiCoSb up to 600~K and carrier concentration 
less than 10$^{21}$ carriers cm$^{-3}$. 
It should be emphasized that experimentally it is often more convenient 
to introduce a small carrier concentration. Thus, a better performance of 
NFNS and TFNS as compared to TiCoSb at smaller doping concentrations could 
be advantageous from experimental perspective. Additionally, the proposed 
values can be further improved by nanostructuring.
Further, in Fig. ~\ref{PF} we have shown the variation of Seebeck coefficient 
and electrical conductivity with temperature for the carrier concentration 
of 4$\times$10$^{20}$ carriers cm$^{-3}$. As we observed from the figure, the 
Seebeck coefficient is dominant for TiCoSb and increases with temperature.
The electrical conductivity, however, dominates in the case  of NFNS/TFNS 
and shows an opposite trend of decreasing values with temperature.

\section{Conclusion}

In conclusion, utilizing first-principles simulations, we have explored
the thermoelectric properties of two disordered half-Heusler alloys
NbFe$_{0.5}$Ni$_{0.5}$Sn and TaFe$_{0.5}$Ni$_{0.5}$Sn. Both systems optimize
in \textit{P$\bar{4}$m2} symmetry and possess a band gap of 0.1 eV. The phonon
dispersions indicate their dynamical stability. Interestingly, the
characteristic gap in acoustic and optical phonons, which imparts high lattice
thermal conductivity to half-Heusler alloys, disappears in both materials.
Likewise, we obtained significantly low lattice thermal conductivity in
NbFe$_{0.5}$Ni$_{0.5}$Sn and TaFe$_{0.5}$Ni$_{0.5}$Sn in comparison to TiCoSb.
We attribute this change in thermal conductivity to the short phonon lifetime
of NbFe$_{0.5}$Ni$_{0.5}$Sn and TaFe$_{0.5}$Ni$_{0.5}$Sn, arising from the
interaction among acoustic and low-lying optical modes. Our simulations on
lattice thermal conductivity with phonon mean free path suggest that,
nanostructuring could bring down the lattice thermal conductivity further.
Electrical transport results reveal that these materials are best suited
in the temperature range 400--600~K and in the carrier concentration
range less than 10$^{21}$ carriers cm$^{-3}$. Within these temperature
and concentration range, we find $\sim$35\% and $\sim$17\% enhancement in the
$ZT$ value for NbFe$_{0.5}$Ni$_{0.5}$Sn and TaFe$_{0.5}$Ni$_{0.5}$Sn, respectively,
as compared to TiCoSb.  We hope that our results would encourage experimental
investigations in these proposed materials.

\begin{acknowledgments}

  M. Z. and C. K. V. are thankful to MHRD and CSIR, respectively, for their
financial assistance. B. K. M. acknowledges the funding support from the
SERB, DST (ECR/2016/001454). The calculations are performed using the High
Performance Computing cluster, Padum, at the Indian Institute of Technology
Delhi.

\end{acknowledgments}

\bibliography{references}

\end{document}